\documentclass[5p]{elsarticle}

\usepackage{color}
\usepackage{booktabs}
\usepackage{tabularx}
\usepackage{import}
\usepackage{amsmath}
\usepackage{amssymb}
\usepackage[T1]{fontenc}
\usepackage{mathtools, cuted}	% strip - multicolumn equation
\usepackage{siunitx}

\usepackage{tikz}
\usepackage{tikzscale}
\usetikzlibrary{calc,fit}
\usetikzlibrary{decorations.pathmorphing}
\usetikzlibrary{decorations.pathreplacing,decorations.markings}

\usepackage{hyperref}

\newcommand{\eq}[1]{Eq.~(\ref{#1})}
\newcommand{\fig}[1]{Fig.~\ref{#1}}
\newcommand{\tab}[1]{Table~\ref{#1}}

\def\Nf{N_\mathrm{f}}
\def\fm{\,{\rm fm}}
\def\MeV{\,{\rm MeV}}
\def\Nl{N_l}

\newcommand{\EQl}{\ensuremath{E_{\bar Ql}}}
\newcommand{\EQs}{\ensuremath{E_{\bar Qs}}}
\newcommand{\rb}[1]{\ensuremath{r^\star_{#1}}}

\journal{Physics Letters B}
\begin{document}
%%%%%%%%%%%%%%%%%%%%%%%%%%%%%%%%%%%%%%%%%%%%%%%%%%%%%%%%%%%%
\begin{frontmatter}
\title{
%$\vphantom{x}$\\[-15mm]\hfill {\normalsize WUB/24-00}\\[5mm]	
	%% Preprint number hack
The quark-mass dependence of the potential energy between static colour 
  sources in the QCD vacuum with light and strange quarks}
%%%%%
\author[1]{John Bulava}
\author[2]{Francesco Knechtli}
\author[2]{Vanessa Koch}
\author[3]{Colin Morningstar}
\author[4]{Michael Peardon}
%%%%%
\affiliation[1]{
   organization={Fakult\"at f\"ur Physik und Astronomie, Institut f\"ur
      Theoretische Physik II, Ruhr-Universit\"at Bochum},
  city={44780 Bochum},
  country={Germany}}
\affiliation[2]{
  organization={Department of Physics, University of Wuppertal}, 
  address={Gau\ss strasse 20, 42119}, 
  country={Germany}}
\affiliation[3]{
  organization={Department of Physics, Carnegie Mellon University},
  citu={Pittsburgh, Pennsylvania 15213},
  country={USA}}
\affiliation[4]{
  organization={School of Mathematics, Trinity College Dublin},
  country={Ireland}}
%%%%%
\begin{abstract}
The low-lying energy spectrum of the static-colour-source-anti-source system 
in a vacuum containing light and strange quarks is computed using lattice QCD 
for a range of different light quark masses. The resulting levels are described
using a simple model Hamiltonian and the parameters in this model are 
extrapolated to the physical light-quark masses. In this framework, 
the QCD string tension is found to be 
  $\sqrt{\sigma}=445(3)_{\rm stat}(6)_{\rm sys}$ MeV. 
\\[-95mm]$\null$\hfill {\normalsize WUB/24-00}\\[87mm]	
	%% Preprint number hack
\end{abstract}
%\begin{keyword}
%\PACS xxx \sep yyy
%\end{keyword}
\end{frontmatter}
%%%%%%%%%%%%%%%%%%%%%%%%%%%%%%%%%%%%%%%%%%%%%%%%%%%%%%%%%%%%
\section{Introduction}
The energy levels of a static quark anti-quark pair $V_n(r)$ as a
function of the interquark distance $r$ probe Quantum Chromodynamics (QCD) 
at all scales. For very small
distances $r\lesssim0.1\,$fm the potential can be computed by perturbation
theory. At large distances the behaviour of the potential depends on the matter
content. In pure-gauge theory at large separations, the potential can be 
modelled by an effective bosonic string. With dynamical quarks, 
the ground-state potential $V_0(r)$ flattens at large $r$ due to the formation 
of a pair of static-light mesons (``light'' refers here to the dynamical 
quarks).  It was noted already in \cite{Sommer:1994fr} that 
``the ground state potential
therefore be called a static quark potential or a static meson potential''.  A
first estimate of the distance where saturation of the ground
state potential sets in was provided in \cite{Alexandrou:1992ti}. Later, a 
model for string breaking as a mixing phenomenon between a string and a
two-meson state was formulated in Ref.~\cite{Drummond:1998ar}, which
predicted for lighter quark mass the energy gap becomes larger and the
width of the mixing region also expands. First observations of string breaking 
were made in the three-dimensional \cite{Philipsen:1998de} and four-dimensional
\cite{Knechtli:1998gf,Knechtli:2000df} SU(2) gauge-Higgs model with scalar 
matter
fields. Evidence for string breaking in QCD with $\Nf=2$ quark flavours was
shown in \cite{Bali:2005fu} for a sea quark mass slightly below the strange
quark mass corresponding to $m_\pi\approx650\,$MeV. String breaking occurred 
for static-source separations of $r_c\approx1.25\,$fm. Our previous work 
\cite{Bulava:2019iut} presented a computation of the three lowest potential 
energy levels in QCD with $\Nf=2+1$ quark flavours at quark masses 
corresponding to $m_\pi\approx280\,$MeV and $m_K\approx460\,$MeV. 
The computation was performed on the N200 gauge ensemble generated by the CLS 
consortium \cite{Bruno:2014jqa}. We found two string breaking distances 
$r_c\approx1.22\,$fm and $r_{c_s}\approx1.29\,$fm corresponding to the 
saturation of the potential levels by two static-light and two static-strange 
meson levels, respectively.

This work extends the analysis of Ref.~\cite{Bulava:2019iut} to study the
quark-mass dependence by including two additional gauge ensembles with 
the same lattice spacing $a\approx0.063\,$fm and with one ensemble at a 
lighter quark mass and one heavier ensemble, covering a range of pion masses
$m_\pi\in[200,340]\,$ MeV. Ref.~\cite{Bulava:2019iut} showed the energy
levels are modelled reliably by a simple three-state Hamiltonian with six
parameters. One model parameter can be interpreted as a string tension. In 
this work we determine the quark-mass dependence of these model parameters. We 
introduce one more parameter in the model to include a curvature term 
proportional to $1/r$ as seen in the Cornell potential \cite{Eichten:1979ms}. 
We extrapolate values of the seven parameters to the physical point following 
a trajectory where the sum of the bare quark masses is kept 
constant and close to its physical value. 

The string tension enters as a parameter into models describing the
fragmentation of partons into hadrons
\cite{Andersson:1983ia,Andersson:1978vj,Andersson:1983jt,Sjostrand:1984ic}. The
so called ``Lund model'' \cite{Andersson_2023} is implemented to describe 
hadronisation in the ``Pythia'' Monte-Carlo event generator
\cite{Sjostrand:2006za}. In such models the string tension corresponds to a
constant energy per unit length of the flux tube containing the colour field of
a quark and anti-quark pair. The result of the present work provides input from
first principles which could be used to refine these models by adding 
information on the excited states, for example.

The paper is organised as follows. The methodology of the lattice calculation, 
closely following the techniques used in our earlier work, is reviewed in 
Sec.~\ref{sec:method}. Section \ref{sec:model} introduces the simple model 
describing the dependence of the lowest energy levels on the 
static-colour-source separation, including the addition of the $1/r$ term. 
Having extracted the model parameters from the data at three values of the 
light quark mass, Section \ref{sec:massdep} establishes their dependence on
the light quark dynamics and extrapolates to the physical quark-mass value. 
This is used to 
present the energy levels in the physical theory as the final result of 
the paper in Fig.~\ref{fig:PotentialPhysical}. Our conclusions are presented in
Section \ref{sec:conclude}. 

\section{Methodology \label{sec:method}}
\begin{table*}[h]
\centering{
\begin{tabular*}{13cm}{ccccllcccc}
label &   $N_{\mathrm{conf}}^{\vphantom{W}}$ & $N_{\mathrm{conf}}^W$ &$t_0/a^2$ 
     &  $N_\mathrm{s}$  &  $N_\mathrm{t}$
  & $m_\pi$[MeV] &   $m_K$[MeV] &  $m_\pi L$\\
\midrule
N203 & 94  & 752  & 5.1433(74) & 48 & 128         & 340 &  440  & 5.4   \\
N200 & 104 & 1664 &  5.1590(76) & 48 & 128        & 280 &  460  & 4.4   \\
D200 & 209 & 1117 &  5.1802(78) & 64 & 128      & 200 &  480  & 4.2             
\\
\end{tabular*}}
\caption{CLS ensembles \cite{Bruno:2014jqa,Bali:2016umi} used in
this work. $N_\mathrm{conf}$ is the number of configurations on which 
fermionic observables were measured. $N_\mathrm{conf}^W$ is the number of 
Wilson loop measurements, binned into $N_\mathrm{conf}$ bins. The third column 
gives the flow scale $t_0$ \cite{Luscher:2010iy} in lattice units. The lattice 
sizes are $T=N_\mathrm{t}a$ in time and $L=N_\mathrm{s}a$ in space. The last 
three columns list the pion mass, the kaon mass and the value of $m_\pi L$ 
from \cite{Bruno:2016plf}.}\label{table:lattice-params} 
\end{table*}

To investigate the dependence of the static energies on quark mass, 
we extend our previous analysis to include 
three ensembles of gauge configurations generated by the CLS consortium
\cite{Bruno:2014jqa,Bali:2016umi}. 
These ensembles include the dynamics of
$\Nf=2+1$ flavours of non-perturbatively O($a$) improved Wilson fermions
\cite{Bulava:2013cta} with a tree-level O($a^2$) improved L\"uscher--Weisz
gauge action \cite{Luscher:1984xn} and are described in 
\tab{table:lattice-params}. 
In addition to ensemble N200 
analysed in Ref.~\cite{Bulava:2019iut}, we consider one ensemble at a smaller 
value of the pion mass, D200 and one at a larger pion mass, N203. 
These ensembles have the same lattice spacing
$a=0.0633(4)(6)\fm$ (corresponding to an inverse bare gauge coupling
$6/g_0^2=\beta=3.55$) \cite{Strassberger:2021tsu}. The quark masses
$m_{\mathrm{up}}=m_{\mathrm{down}}=m_{l}$ and $m_{\mathrm{strange}}=m_{s}$ vary
along a chiral trajectory with constant sum of the bare quark masses. 
The bare
quark masses were chosen such that the chiral trajectory approximately passes
through the physical point \cite{Bruno:2016plf}. 
Since the bare parameters corresponding to the physical point are 
only determined after completing the analysis of the trajectory, some
discrepancy can remain. For the ensembles considered in this work, these 
quark mass mistunings have been computed in \cite{Bruno:2016plf,
Strassberger:2021tsu}. We are confident the effects these mistunings
have on our analysis are small and subsequently neglect them. 
We introduce the quark-mass
parameter, 
\begin{equation}\label{e:x}
\mu_l=\frac{3m_\pi^2}{m_\pi^2+2m_K^2}\,, 
%\mu_l=\frac{3}{2}\frac{m_\pi^2}{\frac{1}{2}m_\pi^2+m_K^2}\,, 
\end{equation}
where $m_\pi, m_K$ are the pion and kaon masses. We take these meson
masses in lattice units from Ref.~\cite{Bruno:2016plf} to compute 
$\mu_l$. To
leading-order in chiral perturbation theory $\mu_l \approx
\frac{3m_{l}}{2m_{l}+m_{s}}$ and so along a trajectory which holds the sum of 
the quark masses $2m_l+m_s$ fixed, 
$\mu_l$ is proportional to the light quark mass $m_{l}$. At the $\Nf=3$
flavour-symmetric point $\mu_l=1$ holds. For physical quark masses and 
correcting for isospin-breaking effects, $m_\pi=134.8\,$MeV, $m_K=494.2\,$MeV 
\cite{Bruno:2016plf} and consequently $\mu_l^{\mbox{\scriptsize phys}}=0.1076$. To mitigate 
against topological freezing, open boundary conditions in time 
\cite{Luscher:2011kk} are used. Reweighting factors 
\cite{Luscher:2012av} 
are included in the analysis of observables to modify the action to the 
appropriate one.  Statistical uncertainties are 
determined by the $\Gamma$-method \cite{Wolff:2003sm,Schaefer:2010hu}. There are no measurable autocorrelations in our data and we do not add a tail to the autocorrelation 
function.  

\begin{figure}[h!]
{\scriptsize
\begin{eqnarray}\label{mixingmat}
%C(\mathbf{r},t)=& 
%  \left(\begin{array}{ccc}
%    \langle\mathcal{O}_{W}(t)            \overline{\mathcal{O}}_{W}(0)\rangle
%  & \langle\mathcal{O}_{B\bar{B}}(t)     \overline{\mathcal{O}}_{W}(0)\rangle
%  & \langle\mathcal{O}_{B_s\bar{B}_s}(t) \overline{\mathcal{O}}_{W}(0)\rangle 
%\\ 
%    \langle\mathcal{O}_{W}(t)            \overline{\mathcal{O}}_{B\bar{B}}(0)\rangle
%  & \langle\mathcal{O}_{B\bar{B}}(t)     \overline{\mathcal{O}}_{B\bar{B}}(0)\rangle
%  & \langle\mathcal{O}_{B_s\bar{B}_s}(t) \overline{\mathcal{O}}_{B\bar{B}}(0)\rangle
%\\
%    \langle\mathcal{O}_{W}(t)            \overline{\mathcal{O}}_{B_s\bar{B}_s}(0)\rangle 
%  & \langle\mathcal{O}_{B\bar{B}}(t)     \overline{\mathcal{O}}_{B_s\bar{B}_s}(0)\rangle 
%  & \langle\mathcal{O}_{B_s\bar{B}_s}(t) \overline{\mathcal{O}}_{B_s\bar{B}_s}(0)\rangle 
%\end{array} \right) \\
%=&
\left(\begin{array}{lll}
\phantom{\sqrt{12}\times} \begin{tikzpicture}[baseline={([yshift=-.1ex]current bounding box.center)}]
\draw[thick] (0,0) -- (0,0.5);
\draw[thick] (0,0.5) -- (0.5,0.5);
\draw[thick] (0.5,0.5) -- (0.5,0);
\draw[thick] (0.5,0) -- (0,0);
\end{tikzpicture}
& \phantom{-}\sqrt{2}\times \begin{tikzpicture}[baseline={([yshift=-.5ex]current bounding box.center)}]
\draw[thick] (0,0) -- (0,0.5);
\draw [decorate, decoration={snake, segment length=0.75mm, amplitude=0.3mm}](0,0.5) -- (0.5,0.5);
\draw[thick] (0.5,0.5) -- (0.5,0);
\draw[thick] (0.5,0) -- (0,0);
\end{tikzpicture} 
& \phantom{-\sqrt{21}\times} \begin{tikzpicture}[baseline={([yshift=-.8ex]current bounding box.center)}]
\draw[thick] (0,0) -- (0,0.5);
\draw [decorate, decoration={snake, segment length=1.05mm, amplitude=0.6mm}](0,0.5) -- (0.5,0.5);
\draw[thick] (0.5,0.5) -- (0.5,0);
\draw[thick] (0.5,0) -- (0,0);
\end{tikzpicture} \\&\\
\phantom{'}\sqrt{2}\times \begin{tikzpicture}[baseline={([yshift=-.5ex]current bounding box.center)}]
\draw[thick] (0,0) -- (0,0.5);
\draw[thick] (0,0.5) -- (0.5,0.5);
\draw [thick](0.5,0.5) -- (0.5,0);
\draw[decorate, decoration={snake, segment length=0.75mm, amplitude=0.3mm}](0.5,0) -- (0,0);
\end{tikzpicture}
& \phantom{--} 2 \times \begin{tikzpicture}[baseline={([yshift=-.8ex]current bounding box.center)}]
\draw [thick](0,0) -- (0,0.5);
\draw [decorate, decoration={snake, segment length=0.75mm, amplitude=0.3mm}](0,0.5) -- (0.5,0.5);
\draw [thick](0.5,0.5) -- (0.5,0);
\draw [decorate, decoration={snake, segment length=0.75mm, amplitude=0.3mm}](0.5,0) -- (0,0);
\end{tikzpicture}+\begin{tikzpicture}[baseline={([yshift=-.8ex]current bounding box.center)}]
\draw [thick](0,0) -- (0,0.5);
\draw (0,0) .. controls (0.1,0.25) .. (0,0.5)[decorate, decoration={snake, segment length=0.75mm, amplitude=0.4mm}];
\draw [thick](0.5,0.5) -- (0.5,0);
\draw (0.5,0) .. controls (0.4,0.25) .. (0.5,0.5)[decorate, decoration={snake, segment length=0.75mm, amplitude=0.4mm}];
\end{tikzpicture} 
& \phantom{12}\sqrt{2}\times \begin{tikzpicture}[baseline={([yshift=-.5ex]current bounding box.center)}]
\draw[thick] (0,0) -- (0,0.5);
\draw [decorate, decoration={snake, segment length=1.05mm, amplitude=0.6mm}](0,0.5) -- (0.5,0.5);
\draw[thick] (0.5,0.5) -- (0.5,0);
\draw[decorate, decoration={snake, segment length=0.75mm, amplitude=0.3mm}](0.5,0) -- (0,0);
\end{tikzpicture} \\&\\
\phantom{\sqrt{12}\times} \begin{tikzpicture}[baseline={([yshift=-.8ex]current bounding box.center)}]
\draw[thick] (0,0) -- (0,0.5);
\draw[thick] (0,0.5) -- (0.5,0.5);
\draw [thick](0.5,0.5) -- (0.5,0);
\draw[decorate, decoration={snake, segment length=1.05mm, amplitude=0.6mm}](0.5,0) -- (0,0);
\end{tikzpicture}
& \phantom{-}\sqrt{2}\times \begin{tikzpicture}[baseline={([yshift=-.8ex]current bounding box.center)}]
\draw[thick] (0,0) -- (0,0.5);
\draw [decorate, decoration={snake, segment length=0.75mm, amplitude=0.3mm}](0,0.5) -- (0.5,0.5);
\draw [thick](0.5,0.5) -- (0.5,0);
\draw[decorate, decoration={snake, segment length=1.05mm, amplitude=0.6mm}](0.5,0) -- (0,0);
\end{tikzpicture}
& \phantom{-\sqrt{12}\times} \begin{tikzpicture}[baseline={([yshift=-.5ex]current bounding box.center)}]
\draw [thick](0,0) -- (0,0.5);
\draw [decorate, decoration={snake, segment length=1.05mm, amplitude=0.6mm}](0,0.5) -- (0.5,0.5);
\draw [thick](0.5,0.5) -- (0.5,0);
\draw [decorate, decoration={snake, segment length=1.05mm, amplitude=0.6mm}](0.5,0) -- (0,0);
\end{tikzpicture}+\begin{tikzpicture}[baseline={([yshift=-.5ex]current bounding box.center)}]
\draw [thick](0,0) -- (0,0.5);
\draw (0,0) .. controls (0.13,0.25) .. (0,0.5)[decorate, decoration={snake, segment length=1.05mm, amplitude=0.6mm}];
\draw [thick](0.5,0.5) -- (0.5,0);
\draw (0.5,0) .. controls (0.37,0.25) .. (0.5,0.5)[decorate, decoration={snake, segment length=1.05mm, amplitude=0.6mm}]; \nonumber
\end{tikzpicture}
\end{array}\right)
\end{eqnarray}
}
\caption{Schematic representation of the string breaking mixing matrix.} 
   \label{figure:mix}
\end{figure}
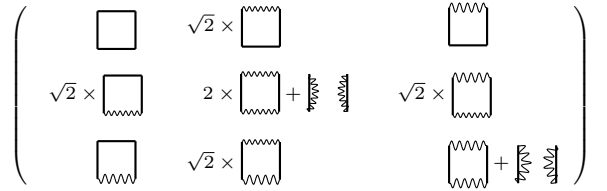

We compute the allowed energies of a system comprising a static quark at 
spatial position $\vec{x}$ and a static anti-quark at position $\vec{y}$ as a 
function of separation $\vec{r}=\vec{y}-\vec{x}$. The calculation involves the 
computation of a matrix of time-correlation functions $C(\vec{r},t)$ between 
states created by a flux tube operator or states consisting of a 
static meson and anti-meson pair with isopin zero. In a theory with degenerate 
dynamical up and down light quarks and a strange-quark, two sets of mesons
are formed by combining a static source with either a light or strange 
quark, giving the static-light or static-strange mesons. $C(\vec{r},t)$, the 
relevant matrix correlation function needed to study this spectrum is shown 
schematically in \fig{figure:mix}.
Explicit expressions for the interpolation operators of string and two-meson states are given in \cite{Bulava:2019iut}.

All gauge links in the interpolating operators are smeared using HYP2 
parameters \cite{Hasenfratz:2001hp,DellaMorte:2003mw,DellaMorte:2005nwx,Grimbach:2008uy,Donnellan:2010mx}
$\alpha_1=1.0, \alpha_2=1.0, \alpha_3=0.5$. For the string operators, 
15 and 20 levels of spatial HYP smearing with parameters $\alpha_2=0.6$,
$\alpha_3=0.3$ \cite{Donnellan:2010mx} are used. To increase spatial
resolution of the energies in the string breaking regions we use on- and
off-axis distances $\vec{r}$. For off-axis displacements, gauge-link paths
which follow the straight line connecting the static source and
anti-source as closely as possible are constructed \cite{Bolder:2000un}, using 
the Bresenham algorithm \cite{Bresenham}. 
Following this closest path significantly enhances the overlap between 
creation operator and physical string states. 

Two distinct computational techniques to compute quark propagators 
(represented by the wiggly lines in \fig{figure:mix}) are needed. 
We refer to propagators starting and terminating on the same time slice 
$t_s=t_{\!f}$ as ``relative'', while quark propagators with $t_s\neq t_{\!f}$ 
are called ``fixed''. Evaluation of quark line diagrams uses the stochastic 
LapH method \cite{Morningstar:2011ka}, based on the distillation 
quark-smearing technique \cite{HadronSpectrum:2009krc}. Distillation projects 
the quark fields on a time-slice into the space spanned by the lowest $N_v$ 
eigenmodes of the three-dimensional gauge-covariant Laplace operator, 
constructed from stout-smeared gauge-links \cite{Morningstar:2003gk} with 
parameters $\rho=0.1$, $n_\rho=36$.  Propagators between distillation spaces 
are estimated using stochastic LapH, with the variance reduced by 
dilution \cite{Foley:2005ac,Bernardson:1993he,Wilcox:1999ab}. Fixed quark
propagators are evaluated on two source times $t_s/a=\{32, 52\}$ using full
time and spin dilution with interlace-8 LapH eigenvector dilution, 
denoted collectively by (TF,SF,LI8) \cite{Morningstar:2011ka}. 
Relative quark propagators are evaluated on all source times $t_s/a\in\{32, 33,
\dots, 95\}$ using interlace-8 time and full spin dilution with interlace-8
LapH eigenvector dilution, labelled (TI8,SF,LI8). A total of $S
\cdot L \cdot N_r \cdot t_s  + S \cdot T \cdot L \cdot N_r $ (fixed+relative)
solutions of the Dirac equation (inversions) are required per gauge
configuration for each quark flavour. Here $N_r$ denotes the number of random
vectors in the space defined by the dilution projectors. For our choice of
dilution schemes the number of inversions per quark flavour is $4 \cdot 8 \cdot
N_r \cdot 2  + 4 \cdot 8 \cdot 8 \cdot N_r $ (fixed+relative). These parameters
are summarised in \tab{table:inv}. Note correlation functions for off-axis 
static source separations are easy to compute using relative and fixed quark 
propagators, unlike the off-axis gauge-link paths.
\begin{table}[h]
  \centering
  \begin{tabular}{c c c c c c c}
    \toprule
     & & & \multicolumn{2}{c}{light} & \multicolumn{2}{c}{strange} \\
     \cmidrule(lr){4-5}
     \cmidrule(lr){6-7}
     & $N_v$ & Type & $N_r$ & $n_\mathrm{inv}$ & $N_r$ & $n_\mathrm{inv}$ \\
     \midrule
    N203 & $192$ & fixed & 5 & 320 & 2 & 128 \\
     & & relative & 2 & 512 & 1 & 256 \\
     \midrule
    N200 & $192$ & fixed & 5 & 320 & 2 & 128 \\
     & & relative & 2 & 512 & 1 & 256 \\
     \midrule
    D200 & $448$ & fixed & 7 & 448 & 2 & 128 \\
     & & relative & 3 & 768 & 1 & 256 \\
     \bottomrule
     \end{tabular}
\caption{The number of Laplacian eigenmodes $N_v$ spanning the distillation 
space and the inversion costs for all ensembles used in this calculation. 
For each type of quark line and flavour, the number of stochastic vectors in 
the space defined by the dilution projectors and the corresponding total number 
of inversions is given. The source times and the dilution schemes employed are 
specified in the text.}\label{table:inv}
\end{table}

\begin{figure}[h!]
{%\small % set default font size for labels, tic marks etc
\subimport{./}{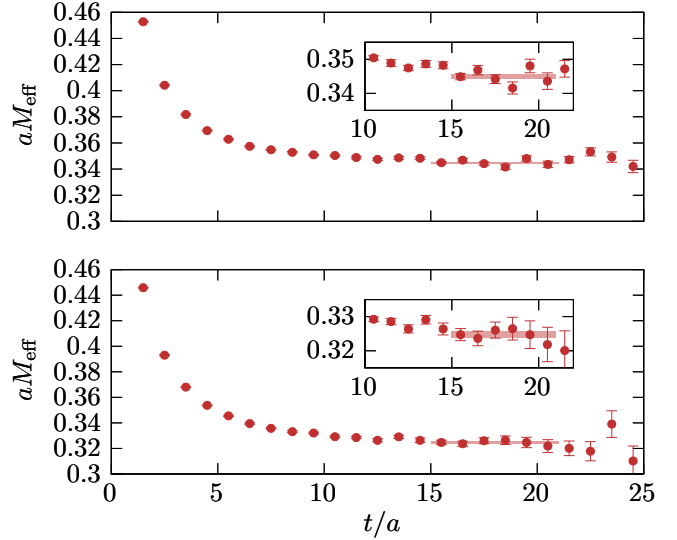}
}
\caption{Effective masses of the static-light (bottom) and 
  static-strange (top) mesons on the ensemble D200. The horizontal bands 
indicate the time intervals used and energies obtained from a single 
exponential fit to the corresponding correlators.}
  \label{figure:Meffstatmesons}
\end{figure}

We start by extracting energies of the static-light meson $E_{\bar{Q}l}$ and 
static-strange mesons $E_{\bar{Q}s}$. 
The thresholds for the avoided level crossings of string breaking are given by
twice these energies.
When an energy level
reaches one of these thresholds, the state formed by a meson anti-meson pair
becomes energetically more favourable. The static-light and static-strange
energies are obtained from single exponential correlated fits to the
time-correlation function $C_{\bar{Q}l}(t)$ and $C_{\bar{Q}s}(t)$ of a
static-light meson and a static-strange meson respectively. For all ensembles
fits are carried out in the time interval $[15,21]$ in lattice units. 
\fig{figure:Meffstatmesons} shows the effective masses $aM_{\rm
eff}(t+a/2)=\ln[C(t)/C(t+a)]$ (points) as a function of $t/a$ for ensemble
D200. The top and bottom panels show the static-strange and 
static-light mesons respectively. Horizontal bands indicate the values of 
$aE_{\bar{Q}l}$ and $aE_{\bar{Q}s}$ extracted from the correlator fits 
extending over the fitted time range.

For each fixed inter-quark separation $r=|\vec{r}|$ we solve a generalized 
eigenvalue problem (GEVP) using the correlation matrix of \fig{figure:mix}
\begin{equation}
        C(t)\,v_n(t,t_0) = \lambda_n(t,t_0)\,C(t_0)\,v_n(t,t_0)\,,
\label{e:gevp}
\end{equation}
to extract the potential energies for $n=0,1,2$ and $t>t_0$. 
We fix $t_0=5a$ for all ensembles. To ensure numerical stability of the 
analysis we first prune \cite{Balog:1999ww,Niedermayer:2000yx,Brett:2019tzr} 
the correlation matrix from size $4\times4$ down to $3\times3$, replacing 
$C(t)$ with 
\begin{equation}\label{e:pruning}
C^{(3)}(t) = 
  \bar{U}^\dagger \bar{C}(t_0)^{-1/2} C(t) \bar{C}(t_0)^{-1/2} \bar{U}. 
\end{equation}
$\bar{U}$ contains the three most significant eigenvectors at $t_0+a$. The bar 
in $\bar{U}$ and $\bar{C}(t_0)$ indicates these matrices are determined on the 
average over gauge configurations. Then we solve the GEVP in \eq{e:gevp} at 
time $t=t_d=10a$, which is fixed for all the ensembles. Using the three 
eigenvectors $v_i(t_0,t_d)$, $i=0,1,2$ we project to a single correlation 
function (``fixed GEVP'') by computing
\begin{equation}\label{e:fixedgevp}
\hat{C}_{ij}(t)=\big(v_i(t_0,t_d),C^{(3)}(t)v_j(t_0,t_d)\big)\,,
\end{equation}
where the parentheses denote the inner product over the components of the 
eigenvectors. The statistical precision of our analysis is improved by 
exploiting the beneficial covariance between $\hat{C}$ and the static-light 
meson correlator $C_{\bar{Q}l}(t)$. 
The correlation function ratios 
\begin{equation}\label{e:ratios}
R_n(t)=\frac{\hat C_{nn}(t)}{C_{\bar{Q}l}^2(t)}\,,
\end{equation}
for $n=0,1,2$ are then computed and fits to a single exponential with time
dependence $\exp(-t(V_n-2\EQl))$ are performed. In these ratios, 
energy levels become renormalised by the
subtraction of twice the divergent static-light energy since the additive 
self-energy contribution of the static-quarks exactly cancels. 
The fit ranges $[t_{\rm min}/a,$ $t_{\rm max}/a]$ % tricky line break
are carefully chosen in each case by 
inspection. Any remaining systematic uncertainty mainly comes from 
$t_{\rm min}/a$ dependence and we choose values of $t_{\rm min}/a$ such that 
these effects are within the limits of statistical precision. 
Finally we set $t_{\rm max}=t_{\rm min}+6a$.  
\begin{figure}[h]
{\small % set default font size for labels, tic marks etc
\subimport{./}{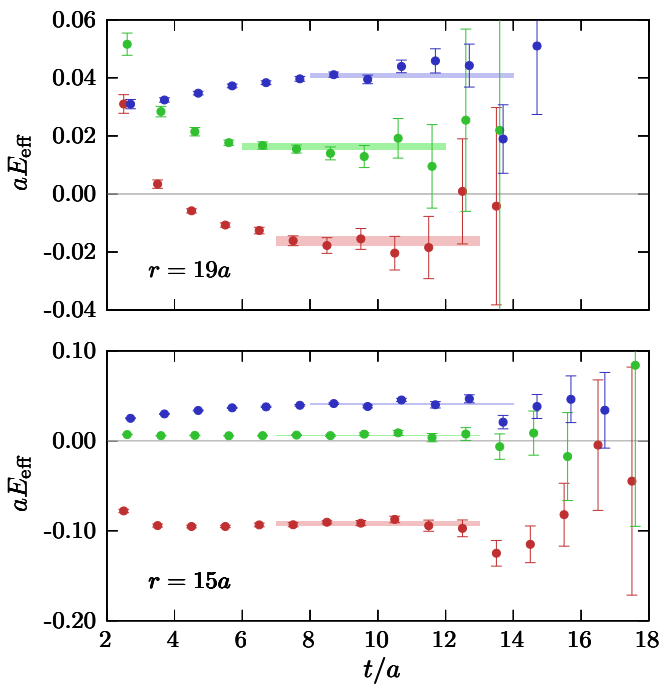}
}
\caption{Examples of effective masses of the ratios \eq{e:ratios} on the
ensemble D200 at two distances as indicated in the plots. The horizontal bands
correspond to the energies $V_n(r)-2E_{\bar{Q}l}$, $n=0,1,2$ extracted from
exponential fits to the ratios and extend over the fitted time
interval.}
\label{figure:Meffpotentials} 
\end{figure}
\fig{figure:Meffpotentials} shows examples of effective masses 
$aE_{\rm eff}(t+a/2)=\ln[R_n(t)/R_n(t+a)]$ for $n=0,1,2$ for two on-axis 
inter-quark separations $r=15a$ and $r=19a$ on the ensemble D200. Note the
effective masses derived from ratios in \eq{e:ratios} can approach their
plateau values from below. The bands indicate the energy
values from the exponential fits of $R_n$ and extend over the fitted time
ranges. The distance $15a$ (bottom plot in
\fig{figure:Meffpotentials}) is the smallest distance down to which we can reliably determine the excited states.
The separation $19a$ (top plot in \fig{figure:Meffpotentials}) is in the
middle of the first string-breaking region where the ground state and first
excited state potential levels would cross.  Notice all three energy levels are
very close at $r=19a$, and yet the GEVP is able to disentangle the three levels
very efficiently.

\fig{figure:VRensembles} shows the three lowest potential energy levels $a(V_n-2\EQl)$ in lattice units as a function of the distance $r/a$ in the region where string breaking occurs for each of the three ensembles described in \tab{table:lattice-params}. Along our trajectory of decreasing light quark mass and increasing strange quark mass the energy gap between the first and second string breaking increases. In order to have a better resolution of the avoided level crossings we include off-axis distances. Some fluctuations in the size of the statistical errors can be seen especially for the off-axis distances. For on-axis distances we profit from self-averaging due to the cubic symmetry.

\section{Model \label{sec:model}}
The model used in Ref.~\cite{Bulava:2019iut} is extended to fit the data 
at smaller separations by the inclusion of a Cornell $1/r$ term. The 
three-state model Hamiltonian becomes
\begin{align}\label{e:3-model7p}
H(r)=
  \left(\begin{array}{rcl}
    \hat{V}(r)& \sqrt{2}g_l & g_s\\  
    \sqrt{2}g_l & \hat E_1 & 0 \\ 
    g_s & 0 &\hat E_2
   \end{array}\right), 
      \hat V(r)=\hat V_0+\sigma r+\gamma/r \,.
\end{align}
After introducing a parameter $\gamma$ for the $1/r$ term,
the new model has seven parameters, labelled
$\{\hat E_1,\hat E_2,g_l,g_s,$ $\sigma,\hat V_0,\gamma\}$. 
The diagonal entries in 
\eq{e:3-model7p}, $\hat V(r)$, $\hat E_1$, $\hat E_2$ are the asymptotic energy 
levels for $r\to\infty$ up to O($r^{-1}$). When there are $\Nl$ degenerate 
light flavours in the mixing matrix elements between string and two-meson states 
of the model Hamiltonian, a flavour factor $\sqrt{\Nl}$ multiplies the mixing 
coefficient $g_l$, in analogy to the corresponding flavour factor in 
\fig{figure:mix}. For $\Nl=2$ the factor $\sqrt{2}$ in the three-state model 
Hamiltonian \eq{e:3-model7p} can be derived from a four-state Hamiltonian with 
non-degenerate up and down quarks by taking the limit of degenerate light 
quarks.
\begin{table}[h]
\centering
\begin{tabular}{cccc}
id &  $n=0$ & $n=1$ & $n=2$ \\
\midrule
N203 & $[4, 24]$  & $[11, 24]$  & $[11, 24]$ \\
N200 & $[4, 24]$  & $[11, 24]$  & $[11, 24]$ \\
D200 & $[4, 27]$  & $[15, 27]$  & $[15, 27]$     
\\
\end{tabular}
\caption{Distance ranges $[r_1(n)/a\,,r_2(n)/a]$ used for the 7-parameter 
model fits to ground ($n=0$), first ($n=1$) and second excited state 
($n=2$).}\label{table:dist-ranges-7p} \end{table}

The dependence of the three eigenvalues $e_n(r),n=0,1,2$ of the model 
Hamiltonian \eq{e:3-model7p} on the distance $r$ is fitted to the three
lowest potential energy levels $V_n(r)$ determined from the data.
These levels are normalised by subtracting twice the energy of
the static-light meson $2 E_{\bar{Q}l}$. Hence the fitted eigenvalues inherit this
normalisation, which removes the divergent self-energy contribution 
from the temporal static-quark lines. The model fit parameters in 
\eq{e:3-model7p} are obtained by minimisation of the correlated $\chi^2$ ,\\
\newpage
\begin{strip}
\begin{align}
\chi^2_{\rm corr} = 
   \sum_{n,m=0}^2 \sum_{r=r_1(n)}^{r_2(n)} \sum_{r'=r_1(m)}^{r_2(m)}  \!
      \Big(V_n(r)-2E_{\bar{Q}l}-e_n(r) \Big) \bar{C}^{-1}  \Big(V_m(r')-2E_{\bar{Q}l}-e_m(r') \Big) , 
  \label{e:chisqcorr}
\end{align}
\end{strip}
where $\bar{C}^{-1}$ is the inverse of the covariance matrix for the potential
levels $V_n$.
The matrix $\bar{C}$ is determined by the $\Gamma$-method.
For each pair of potential levels we compute the squared statistical errors of their sum
and their difference. The covariance between two levels is obtained from the difference
of these two squared errors. In order to obtain a positive matrix we switch off
the autocorrelations between the potential levels when computing $\bar{C}$.
The distance ranges $[r_1(n),r_2(n)]$ used to fit potential levels
$V_n$ on individual ensembles are given in \tab{table:dist-ranges-7p}. The
minimal distance for the ground state $V_0$ is set to $4a$ in order to be
sensitive to the Cornell $1/r$-term. For the first and second excited states, 
$V_1$ and $V_2$ respectively, the minimal distance is chosen so the energy
gap to the ground state is smaller than $2m_\pi$. By this choice we avoid
fitting our excited state levels at too small distances where there could be
unresolved intermediate energy levels in our analysis. For ensembles N200 and 
N203 we set the maximal distance for the fit equal to $L/2$. Potential values 
at distances larger than $L/2$ suffer from finite volume effects and are not 
considered. We impose a cut on the covariance matrix $C$ of the potential 
levels using a singular value decomposition. Eigenvalues of $C$ less than 
$10^{-4}$ times the maximal eigenvalue are removed from the inverse
covariance which we denote by $\bar{C}^{-1}$ in \eq{e:chisqcorr}. We
investigated the dependence of the fit parameters on this cut parameter and
found this a reasonable choice; cuts of $10^{-5}$ or $10^{-3}$ times the
maximal eigenvalues give consistent results.
\begin{figure*}[h]
\begin{centering}
{\small % set default font size for labels, tic marks etc
\subimport{./}{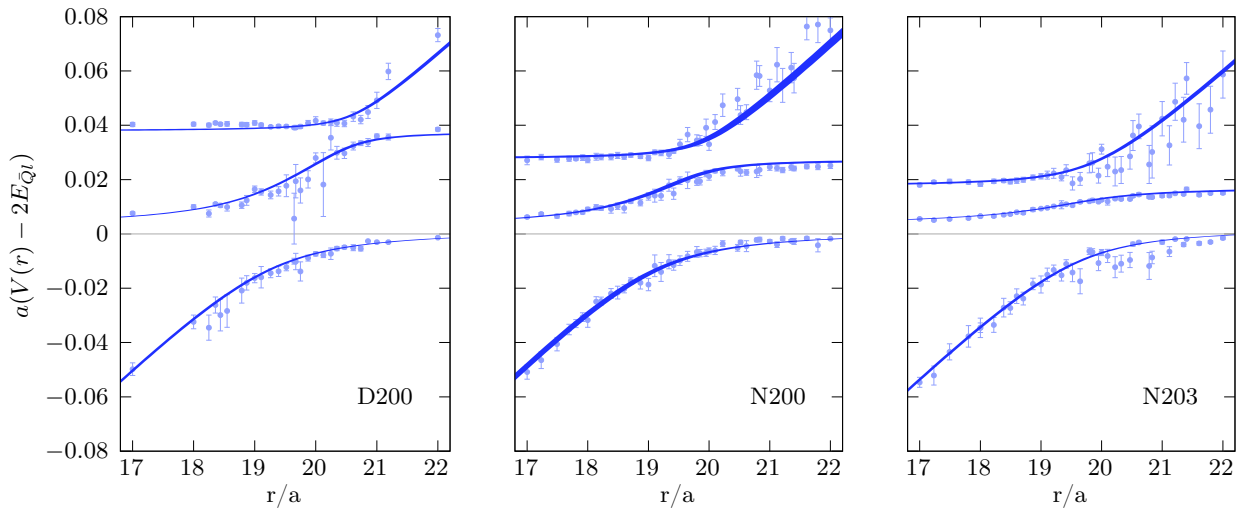}
}
\caption{The energy levels extracted from the GEVP analysis for separations 
in the range $r/a\in[17,22]$ for the three ensembles described in 
  \tab{table:lattice-params}. The data include off-axis separations. The bands 
  indicate the energies extracted from a fit of data to the seven-parameter 
  model of \eq{e:3-model7p} }
\label{figure:VRensembles}   
\end{centering}
\end{figure*}

In \fig{figure:VRensembles} the bands show the seven-parameter model fits to \eq{e:3-model7p} for each of the three ensembles analyzed in this work and including statistical errors. We see that the model describes our data very well in the region where string breaking happens.

\section{Mass dependence and physical point \label{sec:massdep}}
\begin{figure}[h]
{\small % set default font size for labels, tic marks etc}
\subimport{./}{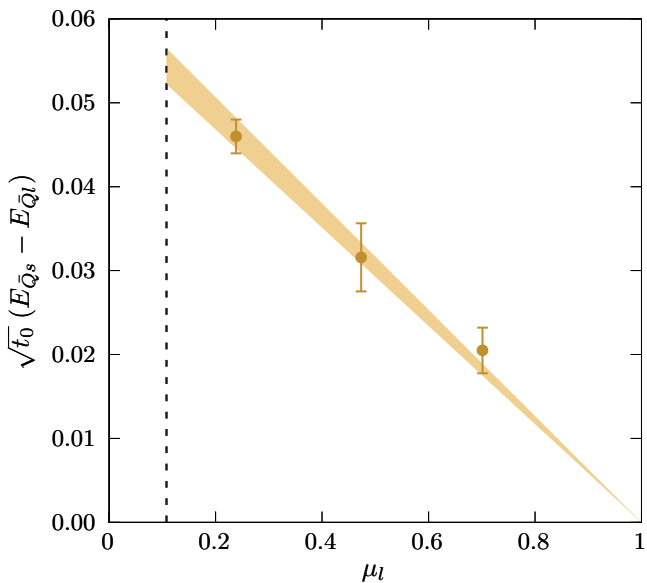}
}
\caption{Dependence of the mass difference between the static-light and 
static-strange meson masses on the light-quark mass parameter 
$\mu_l$ of \eq{e:x}. 
\label{fig:dStatMassdep} } 
\end{figure}

With three ensembles generated with different light and strange quark masses, 
the mass dependence of the model and its parameters can be investigated and an
extrapolation to the physical values of the quark masses attempted. To begin, 
the difference between the static-light and static-strange meson masses was 
determined and a linear extrapolation to the physical light quark mass 
performed. The result of this extrapolation can be seen in 
Fig.~\ref{fig:dStatMassdep}. 
The subtraction removes the divergent static source energy, yielding a
splitting with a well-defined continuum limit.  We performed a fit to a simple
model assuming linear dependence of this splitting on the light-quark mass
parameter, $\mu_l$ away from the three-flavour symmetric point corresponding to
$\mu_l=1$.  The data clearly shows agreement with this simple behaviour.
Extrapolation to $\mu_l^{\mbox{\scriptsize phys}}$ yields
$\sqrt{t_0}(\EQs-\EQl)= 0.054(2)$ which gives 
$\EQs-\EQl=74(3)_{\rm stat}(1)_{\rm sys}\MeV$ using $\sqrt{t_0} = 0.1443(7)(13)$ fm \cite{Strassberger:2021tsu}.  
This compares with the
experimentally determined value of the mass difference, averaged over isospin
for the $B_s$ and $B$ mesons of 87.42(14) MeV\cite{ParticleDataGroup:2022pth}. 
The discrepancy is most likely due to finite heavy-quark-mass effects, 
neglected in this calculation. Other possible effects include 
finite-lattice-spacing artefacts and systematic uncertainty in the light-
and strange-quark mass determinations. 

\begin{figure}[h]
{\small % set default font size for labels, tic marks etc}
\subimport{./}{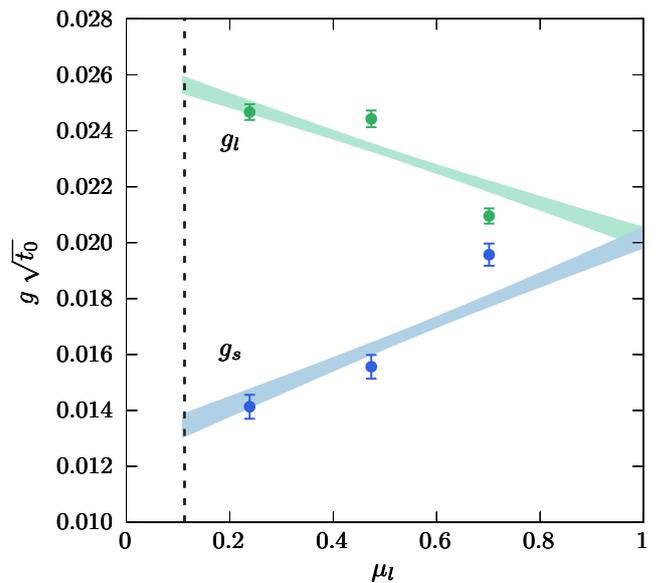}
}
\caption{Dependence of the mixing parameters, $g_l$ and $g_s$ on the 
light-quark mass parameter, $\mu_l$ of \eq{e:x}. 
The vertical line indicates the physical light-quark mass ratio. 
\label{fig:gMassdep} } 
\end{figure}

The two mixing parameters, $g_l$ and $g_s$ of 
\eq{e:3-model7p} determined at the three light-quark values are displayed in 
Fig.~\ref{fig:gMassdep}. 
Without a robust prediction from theory, 
their light-quark mass dependence was again modelled assuming a simple linear 
behaviour, cf. \cite{Grinstein:1996gm} with the two straight-lines constrained to a common
value at the flavour-symmetric point, corresponding to $\mu_l=1$.
For this fit, $\chi^2/n_{\rm df}=16.8$ and some tension between the data and 
the model is seen close to the three-degenerate-flavour point. In the absence 
of a more detailed model for this mass dependence, the phenomenon was not 
investigated further. As seen clearly in this figure however, the two mixing 
parameters, $g_l$ and $g_s$ differ by a factor of almost two for physical 
values of the quark masses.  Assuming linear behaviour
yields extrapolated values of $g_{l,s}$ at the physical light-quark mass 
ratio of 
\begin{align}
   g_l\sqrt{t_0} = 0.0256(3),\quad&\quad 
         g_l = 35.0\,(5)_{\rm stat}(3)_{\rm sys} \MeV
\nonumber\\
   g_s\sqrt{t_0} = 0.0134(4),\quad&\quad 
         g_s = 18.3\,(6)_{\rm stat}(2)_{\rm sys} \MeV. 
\end{align}
%%%%%
\begin{figure}[h]
{\small % set default font size for labels, tic marks etc}
\subimport{./}{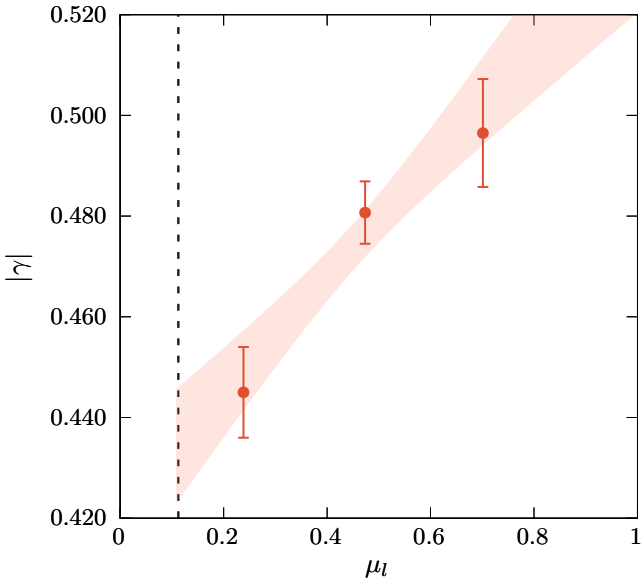}
}
\caption{Light-quark mass-dependence of the coefficient of the dimensionless 
$1/r$ Cornell potential term from fits to the model of \eq{e:3-model7p}. 
The vertical line indicates the physical light-quark mass ratio. 
   \label{fig:gammaMassdep} } 
\end{figure}
%%%%%
\fig{fig:gammaMassdep} shows the light-quark mass dependence of the 
coefficient $\gamma$ weighting the $1/r$ term in $\hat V(r)$ modelling the 
gluonic flux-tube in 
\eq{e:3-model7p}. $|\gamma|$ is plotted for clarity. Significant mass 
dependence in this term is observed and the value extrapolated to the physical 
point is found to be 
\begin{align}
  \gamma^{\rm phys}=-0.434(11). 
\end{align}
The uncertainty is purely statistical as there is no scale dependence on this 
dimensionless parameter. Understanding the mass dependence in detail would
require closer study of the transition from the short-distance behaviour of the
static potential into the string-breaking region and is beyond the scope of
this first study.
%%%%%

\begin{figure}[h]
{\small % set default font size for labels, tic marks etc}
\subimport{./}{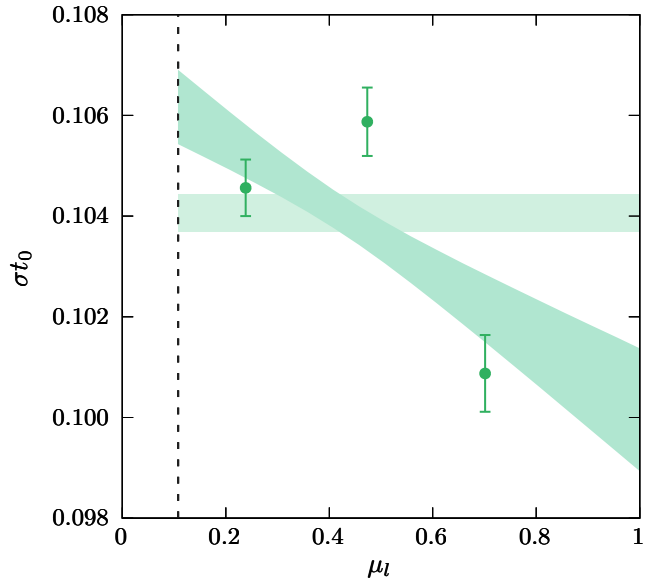}
}
\caption{Dependence of the string tension on the light and strange quark 
masses. $\mu_l$, the dimensionless light quark mass parameter is defined in 
\eq{e:x}.  The two bands indicate extrapolating fits using constant or linear 
dependence.
The physical value of the mass ratio, $\mu_l^{\mbox{\scriptsize phys}}$ is indicated by 
the vertical dashed line.\label{fig:SigmaMassdep} }
\end{figure}

The dependence of the string tension on $\mu_l$ is seen in
Fig.~\ref{fig:SigmaMassdep}, along with two simple models to describe this
behaviour and extrapolate to the physical point.  The simplest model asserts no
dependence of $\sigma$ on $\mu_l$ while the second model assumes linear
dependence.  The change in $\sigma$ over the full range of $\mu_l$ is seen to
be mild and is only a few percent from the three-flavour symmetric theory to
the physical point. The mass dependence is not described very well by either
model and the statistical uncertainties at the physical point from the two fits
do not capture the discrepancy between the models.  Using the linear model to
give a central value at the physical point but assigning a systematic
uncertainty from the difference between the extrapolated values in the two
models yields
\begin{align}
  \sigma t_0 &= 0.1061(7)(20), \quad 
     \sqrt\sigma = 445(3)_{\rm stat}(6)_{\rm sys} \MeV,
  \label{e:sigma}
\end{align}
where the uncertainty quoted includes both statistical and extrapolation 
uncertainties from our determination of $\sigma t_0$ combined in quadrature 
with the scale-setting uncertainty from $\sqrt{t_0} = 0.1443(7)(13)$ fm 
\cite{Strassberger:2021tsu}. 
\tab{tab:covar} shows the correlations between the model parameters after 
extrapolating to the physical quark masses. Some parameters are seen to be
strongly correlated, in particular the three coefficients in $\hat V(r)$ have
high statistical correlations, including a pair with a correlation of $92\%$.
The right table in \tab{tab:covar} lists the model parameters with their statistical errors
after linear extrapolations in $\mu_l$ (best fit) to the physical quark masses.
\begin{table*}
\begin{center}
\begin{tabular}{@{}l|SSSSSScr|S[table-format=+1.5(2)]@{}}
\multicolumn{1}{l}{} 
 &{$\hat E_2$}&{$g_l$}&{$g_s$}&{$\sigma$}&{$\hat V_0$}&{$\gamma$}&\hspace{2.5cm}
   & {Parameter} & {\hspace{-2mm}Best Fit}\\
\cline{2-7} \cline{9-10}
   $\!\!\hat E_1\;$ & 0.12 & 0.23 & -0.02 & -0.01 & -0.01 & 0.10 &
     & $\hat E_1 \sqrt{t_0}$ &  0.0034(3) \\
   $\!\!\hat E_2$ &      & 0.05 &-0.11 & 0.09 & 0.07 &-0.07 & 
     & $\hat E_2 \sqrt{t_0}$ &  0.1005(7) \\
   $\!\!g_l$           &      &      & 0.29 & 0.04 &-0.07 & 0.04 &
     & $g_l \sqrt{t_0}$      &  0.0256(3) \\
   $\!\!g_s$           &      &      &      & 0.06 &-0.13 & 0.13 &
     & $g_s \sqrt{t_0}$      &  0.0134(4) \\
   $\!\!\sigma$        &      &      &      &      &-0.92 & 0.70 & 
     & $\sigma t_0    $      & 0.1061(7) \\
   $\!\!\hat V_0$      &      &      &      &      &      &-0.85 & 
     & $\hat V_0 \sqrt{t_0}$ & -0.835(7) \\
\multicolumn{1}{l}{}   &      &      &      &      &      &      & 
     & $\gamma$              & -0.434(11) 
\end{tabular}
\end{center}
\caption{Summary of the model parameters of 
   \eq{e:3-model7p} after extrapolation to the physical quark masses.
    The left-hand table shows the normalised correlation coefficients between
    all parameters. Their values and statistical uncertainty are given
    in the right table. 
      \label{tab:covar} }
\end{table*}

\begin{figure*}[h]
\begin{centering}
{\small % Reduce the fontsize of the axis number
\subimport{./}{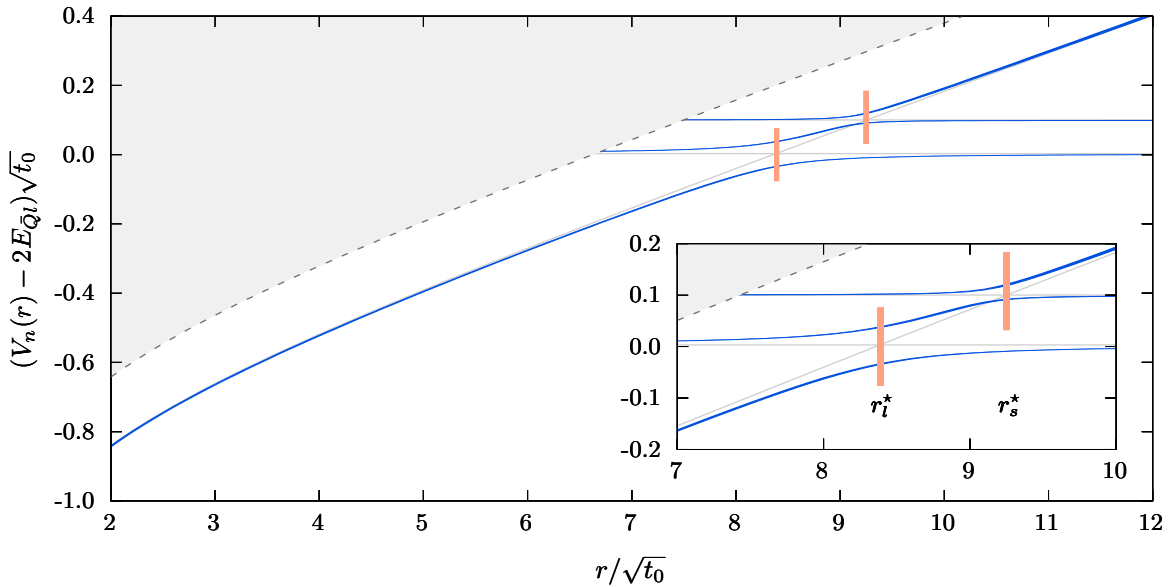}
}
\caption{The energy spectrum of the static source-anti-source system from the
model of \eq {e:3-model7p} with parameters extrapolated in light-quark mass to
the physical point. The grey region denotes the energy regime where two-pion
creation operators would be needed for a robust determination of higher-lying 
states. The breaking distances \rb{l} and \rb{s} computed in the model are 
indicated by the two vertical bands.
    \label{fig:PotentialPhysical}}
\end{centering}
\end{figure*}

Using the model with parameters extrapolated to the physical point, 
the dependence of the energy spectrum of the static source-anti-source system 
on source separation in the presence of physical up, down and 
strange quarks can be computed and is presented in \fig{fig:PotentialPhysical}. 
The zero of the potential is fixed by
subtracting twice the energy of the static-light system, which makes
the ground-state energy at large separations vanish asymptotically. 
The excited states are not displayed at short separations, where the 
basis of operators may not couple efficiently to possible multi-meson states 
and so the determination of the spectrum in this range for this calculation 
is uncertain. 

The string-breaking distances, 
\rb{l} and \rb{s} are indicated by two vertical bands in \fig{fig:PotentialPhysical}. 
These distances are derived from the model by determining the static 
source-sink separation where $\hat V(r)$ is equal to the asymptotic energy of 
two static-light (for \rb{l}) or static-strange (for \rb{s}) mesons. This 
yields
\begin{align}
  \rb{l} &= 8.39(3) \sqrt{t_0}\; =1.211(7)_{\rm stat}(11)_{\rm sys}\;\mbox{fm},
      \nonumber \\
  \rb{s} &= 9.26(2) \sqrt{t_0}\; =1.336(7)_{\rm stat}(12)_{\rm sys}\;\mbox{fm}.
\end{align}
Again, the values quoted in physical units include estimates of the 
uncertainties arising both from limits to the statistical precision of our 
calculation and systematic uncertainty from setting the physical scale. 

\section{Conclusions \label{sec:conclude}}
This paper extends the analysis of Ref.~\cite{Bulava:2019iut} to investigate
the light-quark mass dependence of the potential energy of a system 
of a static colour-anti-colour source pair. The three lowest energy levels in
the resulting spectrum are determined up to the scale where mixing between
states resembling a gluon string and two static-light or static-strange 
mesons is largest. A robust determination is achieved by computing 
correlations within a suitable variational basis of interpolating fields and 
solving the resulting generalised eigenvalue problem. A simple model of the
spectrum is presented, starting from the Cornell potential and allowing mixing 
with the asymptotic static-meson states. 

Using the new data from calculations with different light-quark masses, an
extrapolation of the model parameters to the physical quark mass values is
performed, enabling us to predict the physical potential in QCD with dynamical
light and strange quarks and to give a simple parameterisation of this
potential and its excitations. Perhaps the most studied parameter is the string
tension where the value determined in this work is seen in \eq{e:sigma}. In
comparison, the string tension in pure-gauge theory is in the range
$\sigma t_0\in[0.143,0.159]$.  These values are computed using data on the
deconfining temperature $T_c/\sigma$ compiled in \cite{Necco:2003vh} combined
with $T_cr_0$ from \cite{Necco:2003vh} and $r_0/\sqrt{t_0}$ from
\cite{Knechtli:2017xgy}.  These results are almost 40\% higher than this work
for $\Nf=2+1$ QCD.  A recent determination in $\Nf=2+1+1$ QCD
\cite{Brambilla:2022het} gives $\sqrt{\sigma} =467(7) \MeV$ or $482(7) \MeV$
depending on the type of Cornell fits to the ground state potential used for
distances below $1 \fm$. This shows there are important physical effects in the
string tension generated by the phenomenon of string breaking.

Our data for the three lowest energy levels of the static potential can be
well fitted assuming the simple model in \eq{e:3-model7p}. Other 
phenomenological models \cite{Bruschini:2020voj} could motivate further 
investigations.

\section*{Acknowledgements}
We thank Christoph Hanhart, Paul P\"utz and Tommaso Scirpa for helpful 
discussions of the
mass extrapolations and Tomasz Korzec for help with the data analysis.
We are grateful to Tom DeGrand for feedback. We thank
Ben H\"orz and Graham Moir for their contributions at an early stage. The
authors gratefully acknowledge the Gauss Centre for Supercomputing e.V.
(\url{www.gauss-centre.eu}) for funding this project by providing computing
time on the GCS Supercomputer JUQUEEN\cite{juqueen} at J\"ulich Supercomputing
Centre (JSC). The code for the calculations using the stochastic LapH method is
built on the USQCD QDP++/Chroma library \cite{Edwards:2004sx}. The statistical error analysis has been performed with an internal package of the ALPHA Collaboration. The work is
supported by the German Research Foundation (DFG) research unit FOR5269 “Future
methods for studying confined gluons in QCD”.
C.M. acknowledges support from the U.S. National Science Foundation (NSF) through
award PHY-2209167.
This work was supported by the
STRONG-2020 project, funded by the European Community Horizon 2020 research and
innovation programme under grant agreement 824093. 
\newpage
%%%%%%%%%%%%%%%%%%%%%%%%%%%%%%%%%%%%%%%%%%%%%%%%%%%%%%%%%%%%
\bibliographystyle{elsarticle-num} 
\bibliography{string}
\end{document}